\documentstyle[11pt,aaspp4,flushrt]{article} 

\begin{document}

\def\spose#1{\hbox to 0pt{#1\hss}}
\def\lta{\mathrel{\spose{\lower 3pt\hbox{$\mathchar"218$}}
     \raise 2.0pt\hbox{$\mathchar"13C$}}}
\def\gta{\mathrel{\spose{\lower 3pt\hbox{$\mathchar"218$}}
     \raise 2.0pt\hbox{$\mathchar"13E$}}}
\def\Msun{{\rm M}_\odot}
\def\msun{{\rm M}_\odot}
\def\Rsun{{\rm R}_\odot}
\def\Lsun{{\rm L}_\odot}
\def\half{{1\over2}}
\def\RL{R_{\rm L}}
\def\zs{\zeta_{s}}
\def\zR{\zeta_{\rm R}}
\def\dJJ{{\dot J\over J}}
\def\dMM{{\dot M_2\over M_2}}
\def\tKH{t_{\rm KH}}
\def\eck#1{\left\lbrack #1 \right\rbrack}
\def\rund#1{\left( #1 \right)}
\def\wave#1{\left\lbrace #1 \right\rbrace}
\def\dd{{\rm d}}


\title{TRANSIENTS AMONG BINARIES WITH EVOLVED LOW--MASS COMPANIONS}
\author{A. R. King, \altaffilmark{1}, J. Frank, \altaffilmark{2}
U. Kolb \altaffilmark{1} and H. Ritter \altaffilmark{3}}

\altaffiltext{1} {Astronomy Group, University of Leicester, 
Leicester LE1 7RH, U.K. (ark@star.le.ac.uk, uck@star.le.ac.uk)}

\altaffiltext{2} {Department of Physics and Astronomy, Louisiana State
University,   Baton Rouge, LA 70803-4001, USA
       (frank@rouge.phys.lsu.edu)}

\altaffiltext{3} {Max--Planck--Institut f\"ur Astrophysik, 
Karl--Schwarzschild--Str. 1, D~85740 Garching, Germany
(hsr@MPA-Garching.MPG.DE)}

\begin{abstract}
We show that stable disk accretion should be 
very rare among low--mass X--ray
binaries and cataclysmic variables whose evolution is driven by the
nuclear expansion of the secondary star
on the first giant branch.
Stable accretion is confined to neutron--star 
systems where the secondary is still 
relatively massive, and some supersoft
white dwarf accretors. All other systems, including
all black--hole systems, appear as soft X--ray transients or dwarf novae.
All long--period neutron--star systems become transient well before
most of the envelope mass is transferred, and remain transient until
envelope exhaustion. This complicates attempts to compare the numbers of 
millisecond pulsars in the Galactic disk with their LMXB progenitors, and
also
means that the pulsar spin rates are fixed in systems which are transient 
rather than steady, contrary to common assumption.
The long--period persistent sources Sco X--2, LMC X--2, Cyg X--2 
and V395 Car must have minimum companion masses
$M_2 \gta 0.75\msun$ if they contain 
neutron stars, and still larger $M_2$ if they
contain black holes. The neutron--star
transient GRO J1744-2844 must have $M_2 \lta 0.87\msun$.
The existence
of any steady sources at all at long periods supports the ideas that
(a) the accretion disks in many, if not all, LMXBs are strongly 
irradiated by the central source, and (b) mass transfer is thermally 
unstable in long--period supersoft X--ray sources.
\end{abstract}


\section{INTRODUCTION}
In a recent paper King, Kolb \& Burderi (1996; hereafter KKB) considered
the question of which low--mass X--ray binaries (LMXBs) should be
transient according to the disk instability model. They 
concentrated mainly on systems with short orbital periods $P \lta 2$ d,
showing in particular that neutron--star systems required secondaries
which were somewhat nuclear--evolved before mass transfer began if they 
were to be transient, while this requirement 
is much weaker for black--hole
systems, so that most are likely to be transient at these periods. The
neutron--star result implies a significant constraint on LMXB formation,
which is discussed further by King \& Kolb (1997).

KKB also briefly discussed the occurrence of transients in longer--period
LMXBs whose evolution is driven by the nuclear expansion of an evolved
companion. They concluded that almost all of these systems should be
transient (their equations 6 and 7). While this conclusion is essentially
correct and in agreement with the observational data, the 
need for a more exact discussion
of long--period transients has been thrown into sharper relief
by the recent work of King et al. (1996). This shows that
LMXBs with evolved companions are potentially subject to a violent
instability arising from irradiation of the companion by the accreting
primary. If unquenched this instability would lead to super--Eddington
accretion rates and catastrophic evolution through a common--envelope
phase to an ultrashort--period binary. Observed LMXBs with evolved
companions must therefore be protected against the instability. King
et al. (1996) identify two ways in which this can happen: first, the
companion may be shielded from the irradiation by an extensive accretion
disk corona or some similar structure. Second, and more commonly, the
instability is quenched by the short duty cycles observed in soft X--ray 
transients.

Accordingly we examine here
in more detail the possibility of transient 
behaviour in such systems. In particular we allow for the higher mass
transfer rates implied by mass ratios close to unity, something neglected
by KKB, but which can suppress transient behaviour. 
We assume throughout that the orbital periods are long enough ($P \gta
1 - 2$~d) that nuclear--driven evolution always
dominates over angular momentum loss by magnetic braking (cf KKB and 
Pylyser \& Savonije 1988a, b). 
We restrict our discussion to
systems with a secondary mass $\lta 2 \msun$ where the
giant donor has a degenerate helium core and is well--established on
the first giant branch.
Hence our considerations (and those of KKB) do not apply to
the black--hole transient GRO J1655-40. 
The optical observations of 
Orosz \& Bailyn (1996) show that the companion star is not one of the
low--mass giants considered here but is instead in a very
short--lived evolutionary phase (crossing the Hertzsprung gap). Evidently 
this evolution does allow for phases of rather slow radius expansion,
giving the low mass transfer rates required for transient behaviour,
but a full evolutionary calculation is needed to check this.

\section{CONDITION FOR A DISK INSTABILITY IN LMXBs}

In this section we essentially
follow the arguments of KKB and van~Paradijs (1996), with a few
refinements which we shall discuss explicitly.
The condition for a disk instability can be summarized  
as the requirement that some part of the
disk should have a temperature below the hydrogen ionization temperature
$T_{\rm H} \sim 6500$ K. If a steady state exists in which all of the
disk is above this temperature the instability will be suppressed.
Since the effective temperature $T_{\rm eff}(R)$ in a steady--state
accretion disk decreases outwards 
with distance $R$ from the central accretor  
as $R^{-3/4}$ (cf eqn. 9 below)
this suppression is hardest to achieve at the outer disk
edge $R_d$, and the condition $T_{\rm eff}(R_d) > T_{\rm H}$ defines
a minimum accretion rate $\dot M$ above which the disk will be steady.
However in LMXBs, van~Paradijs (1996) has emphasized the
dominant importance of irradiation of the disk surface by the central 
accreting source, which produces an effective temperature 
$T_{\rm irr}$ given by
\begin{equation}
T_{\rm irr}^4 = {\eta\dot M c^2(1-\beta)\over 4\pi\sigma R^2}{H\over R}
                \rund{{{\rm d}\ln H\over {\rm d}\ln R}-1}\; .
\label{e2.1}
\end{equation}
Here $\sigma$ is the Stefan--Boltzmann constant, $\beta$ the albedo 
of the disk surface, and $H(R)$ the local disk scale height. The factor
$\eta$ measures the efficiency of conversion of potential energy 
into irradiating flux. This
necessarily absorbs a number of unknown factors of order unity 
which arise from e.g.\ limb darkening, the spectral 
energy distribution of the emergent flux, or the fact that the effective 
mass--radius relation of the accretor is not exactly known. Hence we do 
not distinguish between the neutron star and the black hole case and
set $\eta = 0.11$ as in KKB. Any 
contribution to $\eta$ from nuclear burning of the accreted matter is 
negligible ($<0.007$) for a neutron star, but need not be for a white 
dwarf.  
The last pair of factors on the rhs are typically $\propto H/R \sim$
constant (see below). They represent $\sin \alpha$, where 
$\alpha$ is the very shallow angle between the incident radiation from
the central point source and the disk surface, assuming small
$\alpha$, $H/R$ and ${\rm d} H/{\rm d} R$.  

From (\ref{e2.1}) we see that $T_{\rm irr}$ 
decreases only as $R^{-1/2}$, and so is likely to exceed $T_{\rm eff}$
at $R_d$. Accordingly we define a new (lower) critical accretion rate
$\dot M_{\rm crit,\ irr}$ by the requirement 
\begin{equation}
T_{\rm irr}(R_d) = T_{\rm H}\; .
\label{e2.2}
\end{equation}
To evaluate $\dot M_{\rm crit,\ irr}$ we assume that $R_d$ is about
70\% of the primary's Roche lobe radius $R_1$. Here we make the first 
refinement: in place of the Paczy\'nski (1971) formula for the latter
quantity, which introduces an awkward logarithmic dependence, we use 
instead the fact that for mass ratios $0.03 < M_2/M_1 < 1$ the ratio
of primary and secondary Roche lobes can be approximated to better than
5\% by
\begin{equation}
{R_1\over R_2} = \rund{{M_1\over M_2}}^{0.45}\; .
\label{e2.3}
\end{equation}
We now assume that $R_2$ is equal to the thermal--equilibrium radius
$R_e$ 
of a (sub)giant of core mass $M_c$, using the approximate expression
\begin{equation}
R_e(M_c) = 12.55\Rsun \biggl({M_c\over 0.25\Msun}\biggr)^{5.1}\; ,
\label{e2.4}
\end{equation}
(Webbink, Rappaport and Savonije 1983; King 1988). 
Combining (\ref{e2.1} -- \ref{e2.4}) we find 
\begin{equation}
\dot M_{\rm crit,\ irr} \simeq 3.81\times 10^{-9}
   \rund{{M_1\over M_2}}^{0.9}\rund{{M_c\over 0.25\msun}}^{10.2}
 \rund{{{\rm d}\ln H\over {\rm d}\ln R} - 1}^{-1}\; \msun\; {\rm yr}^{-1}
\: ,
\label{e2.5}
\end{equation}
where we have used $\beta = 0.9$ and $H/R = 0.2$, as in KKB.
As long as the mass transfer rate $-\dot M_2$ is above the limit
(\ref{e2.5}) disk
accretion is always stable. Hence $-\dot M_2 < \dot M_{\rm crit,\ irr}$
is a necessary condition for a disk instability to occur.

We can compare (\ref{e2.5}) with the mass transfer rate driven by nuclear
evolution
\begin{equation}
-\dot M_2 = 9.74\times 10^{-9}(\zeta_e - \zR)^{-1}\rund{{M_2\over \msun}}
     \rund{{M_c\over 0.25\msun}}^{7.1}\; \msun\; {\rm yr}^{-1}\ ;
\label{e2.6}
\end{equation}
obtained by requiring the stellar radius to move in step with the Roche 
lobe  (see Webbink et al. 1983, or King 1988); 
here $\zeta_e$ is the thermal equilibrium mass--radius index (taken
as zero in the two papers quoted) and $\zR
\simeq 2 M_2/M_1 - 5/3$
gives the corresponding Roche lobe motion. This is an 
improvement on equation (7) of KKB, which implicitly assumed $M_2<<M_1$
and thus took the factor $(\zeta_e-\zR)^{-1}$ as constant. As we shall 
see, mass ratios $M_2/M_1 \gta 0.7$ can raise $-\dot M_2$ significantly 
towards $\dot M_{\rm crit,\ irr}$.

We see from (\ref{e2.5}, \ref{e2.6}) that $\dot M_{\rm crit,\ irr}$
increases faster with $M_c$ than $-\dot M_2$. Thus disk accretion is
unstable above a critical core mass 
\begin{equation}
M_{c, {\rm crit}} = 0.338\msun(\zeta_e-\zR)^{-0.32}
 \rund{{M_2\over \msun}}^{0.32}\rund{{M_1\over M_2}}^{-0.29}
 \rund{{{\rm d}\ln H\over {\rm d}\ln R} - 1}^{0.32}\; .
\label{e2.7}
\end{equation}
The critical core mass is plotted against $M_2$ in Figure~1, with
$\zeta_e = -0.2$ (King et al. 1996) and $H \propto R^{9/7}, R^{9/8}$
(KKB) for a neutron star ($M_1=1.4\msun$) and a black hole 
($M_1=10\msun$). We also show 
the evolutionary tracks $M_c \sim M_2^{-0.288}M_1^{-0.392}$ 
(cf King 1988).
The companion star specified by a pair of values $(M_2, M_c)$ on
Fig.~1 would fill its Roche lobe in
a binary of orbital period $P$ given by
\begin{equation}
M_c = 0.23\msun\rund{{P\over 10\ {\rm d}}}^{0.13}
\rund{{M_2\over \msun}}^{0.065}\; .
\label{e2.8}
\end{equation}
As the dependence on $M_2$ is extremely weak we can regard $P$ as given 
purely by $M_c$; we mark the corresponding $P$--values on the right--hand
scale in Fig.~1.
Mass transfer in systems with a $1.4\msun$ neutron star and a giant
donor with mass $\ga 1\msun$ is either dynamically or thermally
unstable; such systems 
would not appear as LMXBs (cf.\ Kalogera \& Webbink 1996). There is no
such limit for systems with more massive black hole accretors, so that
the lower 
panel of Fig.~1 could easily be extended up to $M_2\simeq2\msun$ by
obvious extrapolation of the curves shown. 

The same donor mass limits apply to systems with asymptotic
giant--branch donor stars.
These binaries have very long orbital periods (several years) and
a much more complicated mass loss history; the mass transfer rate
is highly modulated by thermal pulses and the wind mass loss rate is
substantial (e.g.\ Pastetter \& Ritter 1989).

It is clear from Fig.~1 that all long--period black--hole LMXBs 
with donors on the first giant branch 
must be transient, as observed.
We also see that in LMXBs with a neutron star primary, the disk can
be stable if the companion mass is high enough, so that the system is
close to thermally unstable mass transfer ($(\zeta_e -\zR) 
\rightarrow 0$). 
These two conclusions imply stringent constraints on the known
persistent LMXBs with long orbital periods. These are currently
Sco X--2 ($P=13.94$~d, Southwell et al. 1996) 
LMC X--2 ($P=12.54$~d), Cyg X--2 ($P = 9.84$~d), and V395 Car = 
2S0921-630 ($P=9.01$~d, e.g. Mason et al. 1987, although the claimed
periodicity is disputed by Krzeminski \& Kubiak 1991). As X--ray
irradiation of both the accretion disc and the companion star must
dominate the optical emission the precise nature of the companions is
not easily deduced from observation. However, we know that they
must be evolved stars, since they
must fill the Roche lobe in a relatively wide binary. Thus 
the low--mass subgiant evolution considered here gives the lowest possible
companion mass. In this case Figure~1 requires minimum companion masses  
$M_2 \gta 0.75\msun$ for all four systems.
This latter value is consistent with Mason et al.'s
mass function for V395 Car. 
If the secondary masses are close to this lower limit, 
all four systems
must contain neutron stars rather than black holes (this is already
suspected for Cyg X--2 as it has shown evidence of X--ray bursts). 
In GRO J1744-2844, the only known long--period neutron--star transient 
with accurately determined orbital period ($P=11.88$~d), 
the small mass function $1.36\times 10^{-4}\msun$ found
by Finger et al. (1996) and the low eccentricity strongly point to an
evolved low--mass companion.
From Fig.~1 we find a maximum secondary mass 
$M_2\lta 0.75\msun$ for the slope $9/7$ ($0.87\msun$ for $9/8$).
With $M_1=1.4\msun$ and the above
mass function this requires inclinations $i \gta 7^{\circ}$ ($9/7$)
and $\gta 6^{\circ}$ ($9/8$), respectively. 

Figure~1 shows that all low--mass neutron--star systems
ultimately become transient as $M_2$ is reduced. As is well 
known, the endpoint of the binary evolution of such systems (when 
$M_2=M_c$) is a millisecond radio pulsar in a wide 
($\sim 100$ d) orbit with a low--mass white dwarf, the remnant core of
the (sub)giant companion after envelope exhaustion. Our result means that
the spin rate of the pulsar at the end of the mass--transfer phase is
determined by this transient phase, rather than in a system transferring
mass steadily at the rate $-\dot M_2$, as is usually assumed. Combined
with the prediction that all black--hole systems are transient, 
the prevalence of transients among neutron--star systems
agrees with the fact that persistent systems are rather rare among 
long--period LMXBs (e.g. van~Paradijs 1995). 
 
From (\ref{e2.6}) we see that even with the most extreme assumptions
($M_2 \approx M_c \approx 0.15\msun$, 
corresponding to short orbital periods
$P \sim$ few days) the mass transfer rate in LMXBs with
evolved secondaries is 
$\gta 10^{-10}\msun\; {\rm yr}^{-1}$. In transients essentially all of 
this
mass must be accreted during outbursts. With typical transient duty cycles
$\lta 10^{-2}$ it follows 
that the outburst accretion rates must be at least
$10^{-8}\msun\; {\rm yr}^{-1}$, i.e. at or above the Eddington limit 
for a $1.4\msun$ neutron star. For larger $M_c$ (longer orbital periods)
or shorter duty cycles the rates are still higher.
These predictions agree with observations of transient outbursts
(e.g. V404 Cyg had a peak outburst luminosity 
$\sim 10^{39}$ erg s$^{-1}$, cf Tanaka \& Lewin 1995; King 1993).

\section{STABILITY OF UNIRRADIATED DISKS}

The arguments above apply to disks where irradiation is
dominant in determining the surface temperature of the disk.
If irradiation does not act upon the disk surface, as e.g. in CVs,
we must replace (\ref{e2.1}) by the standard effective temperature
for a steady disk powered by local viscous dissipation (e.g. Frank, 
King \& Raine 1992)
\begin{equation}
T_{\rm eff}^4 = {3GM_1\dot M\over 8\pi\sigma R^3}\; .
\label{e3.1}
\end{equation}
For mass transfer driven by nuclear evolution we follow the same steps
as in Section 2,  
and find a new critical core mass (independent of the nature of the 
primary)
\begin{equation}
M_{c, {\rm crit}} = 0.097(\zeta_e-\zR)^{-0.12}
 \rund{{M_1\over \msun}}^{0.29}\rund{{M_2\over \msun}}^{-0.04}\; .
\label{e3.2}
\end{equation}

The lowest
value of $M_c$ possible in the nuclear--driven evolution we are 
considering is $\sim 0.15\msun$. Thus except for thermally unstable mass
transfer ($(\zeta_e-\zR) \longrightarrow 0$), the critical value of
$M_c$ given by (\ref{e3.2}) is so small that we expect unirradiated disk
accretion to be unstable in any system with an evolved low--mass companion.
 
Since white dwarfs cannot exceed the Chandrasekhar limit, mass transfer
stability demands low--mass companions.
We conclude that all long--period CVs should be dwarf novae, while the
supersoft sources (thought to be undergoing thermal--timescale mass 
transfer,
cf.\ e.g.\ DiStefano \& Nelson 1996) 
can have stable disks. The only two known CVs at long enough
orbital periods for our considerations to apply, namely V1017 Sgr
($P=5.714$~d) and GK Per ($P=1.997$~d) are both observed to have
dwarf nova outbursts. For LMXBs we would conclude that unirradiated
steady disks are impossible for long orbital periods. The fact that
persistent systems are nevertheless observed at such periods
(see Section 2) shows that
irradiation must effectively determine the surface disk temperatures in
these systems, and presumably most or all LMXBs.

\section{CONCLUSIONS}

We have shown that stable disk accretion is very rare among 
LMXBs and CVs with evolved companions
on the first giant branch.
Apart from a few systems (such as Sco X--1, cf KKB) 
with short enough orbital periods that magnetic 
braking dominates nuclear evolution, it is confined (at best) to
the early part of neutron--star binary evolution, i.e.
systems with quite large secondary masses $M_2$, and some of the 
supersoft 
white--dwarf X--ray sources. 
All long--period CVs and black--hole systems 
with donors on the first giant branch,
as well as most neutron--star systems at these periods, 
have unstable disks and appear as dwarf novae or soft X--ray
transients. These conclusions are in excellent agreement with 
observation, and strongly support the view (van~Paradijs 1996)
that X--ray irradiation determines the disk temperatures in most if not
all LMXBs. Since the neutron--star transients will ultimately produce
millisecond pulsars it is clear that any claims of possible discrepancies
between the numbers of the latter and those of their LMXB progenitors 
(at least in the Galactic disk) must deal with the complex selection 
effects involved in the discovery of transients. Further,
all millisecond pulsars in long--period detached binaries must have
descended from transients, and this must be considered in trying to
work out their spin histories.

Even when the stringent conditions (cf Fig. 1 and Section 2) for 
steady disk accretion are met, persistent LMXBs still have somehow
to suppress the violent instability resulting from irradiation of an
evolved companion (King et al. 1996). Since one cannot invoke transient
behaviour (by definition), these systems must shield the companion from
this radiation in some way, probably by means of an accretion disk
corona. We note that the presence of such coronae is inferred in 
Cyg X--2 and V395 Car, and that
indeed none of the persistent LMXBs where the 
period is known or suspected 
to be long is observed to have a clear star--by--star
eclipse. This supports the idea that accretion disk coronae do not merely
make it impossible to observe eclipsing sources, but are essential
to the survival of the systems themselves.

JF thanks the Max--Planck--Institut f\"ur Astrophysik for warm
hospitality during a productive stay in June--August 1996. This
work was partially supported by the U.K. Science and Engineering
Research Council (now PPARC) and by NASA grant NAG5-2777 to LSU. 
ARK acknowledges support as a PPARC Senior Fellow, and the warm hospitality
of the MPA.
We thank the referee, Deepto Chakrabarty, for helpful comments.

{}

\clearpage

\begin{figure}
\caption{ 
({\it a}) The critical core mass vs total secondary mass for an LMXB  
with a $1.4\msun$ neutron--star primary (see text for assumptions). 
The two solid curves give the boundaries between persistent and transient
behaviour for disk scale 
heights varying as $H \propto R^{9/7}, R^{9/8}$ respectively. 
The scale on the right--hand axis gives the
approximate binary period $P$ at which the companion would fill its Roche 
lobe (cf equation (8)
of the text).
Also shown (dashed) is the 
track in the $M_c - M_2$ plane followed by a binary driven by nuclear
expansion of the companion. This star has core and total
masses $M_c(0) = 0.23\msun, M_2(0)=1.5\msun$ at the start of mass 
transfer.  The positions of Sco~X--2, 
LMC X--2, Cyg X--2 and V395 Car corresponding to the minimum
allowed secondary masses for persistent behaviour
are all very close to the filled circle.
The maximum secondary mass for the neutron--star transient GRO
J1744-2844 is again close to $0.75 \msun$ (for slope $9/7$), or close
to $0.87 \msun$ (for slope $9/8$).
({\it b}) As ({\it a}), but now for a $10\msun$ black
hole primary. All these systems are transient.
}
\end{figure}

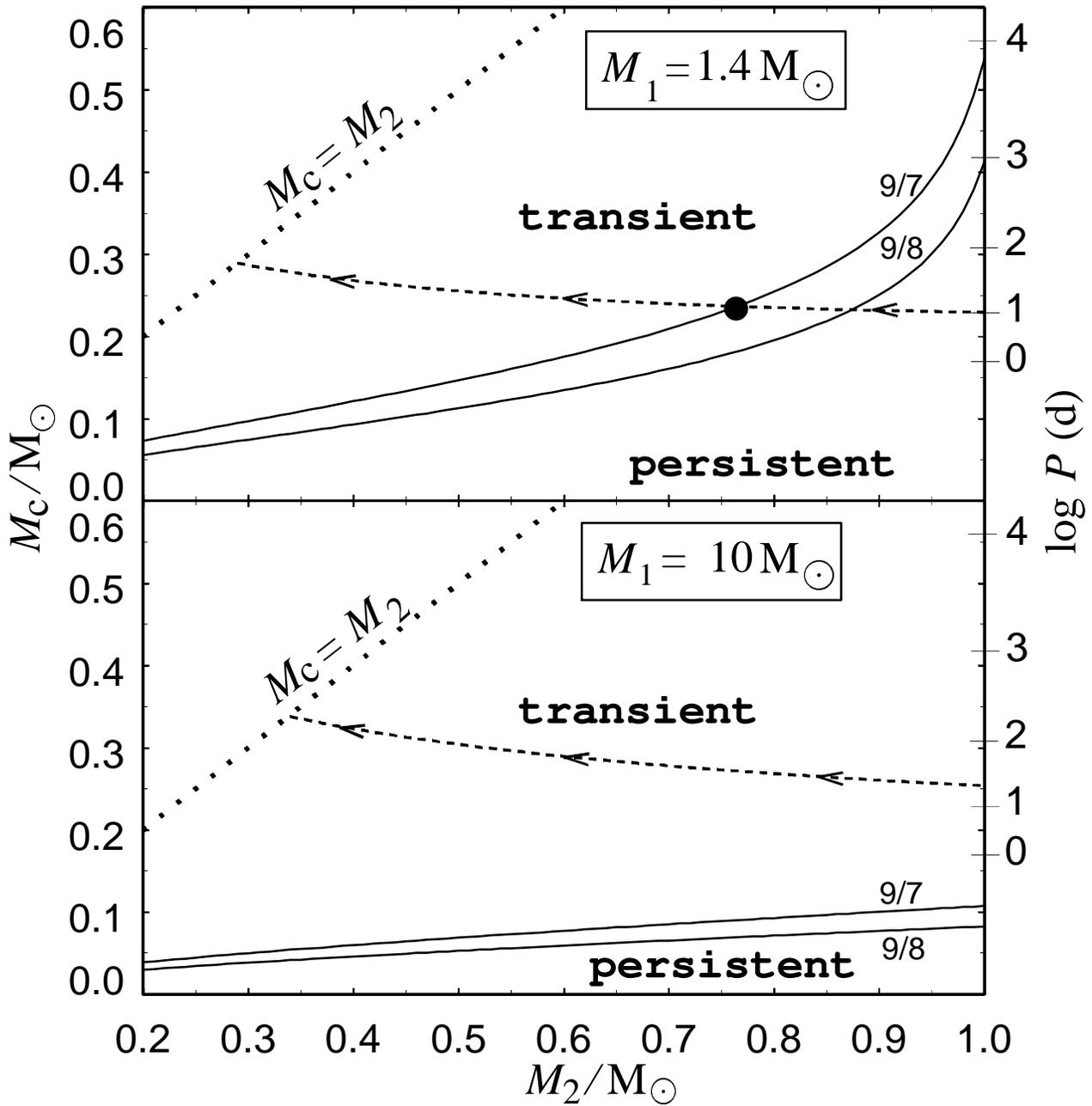
\end{document}